\documentclass[reprint,
aps,pra,
amsmath,amssymb,
superscriptaddress,
]{revtex4-2}
\usepackage{graphicx}% Include figure files
\usepackage{dcolumn}% Align table columns on decimal point
\usepackage{bm}% bold math
\usepackage{siunitx}% SI unit package SI{value}{units}
\usepackage[colorlinks=true,allcolors=blue,bookmarks=false,pdfusetitle]{hyperref}

\begin{document}

\title{Quantum Fourier transform on photonic qubits using cavity QED}
\author{Yu Shi}
 \email{shiyu@terpmail.umd.edu}
 \affiliation{Department of Electrical and Computer Engineering and Institute for Research in Electronics and Applied Physics, University of Maryland, College Park, Maryland 20742, USA}
\author{Edo Waks}
 \email{edowaks@umd.edu}
 \affiliation{Department of Electrical and Computer Engineering and Institute for Research in Electronics and Applied Physics, University of Maryland, College Park, Maryland 20742, USA}
 \affiliation{Joint Quantum Institute, University of Maryland, College Park, Maryland 20742, USA}
 \affiliation{Department of Physics, University of Maryland, College Park, Maryland 20742, USA}

\date{\today}

\begin{abstract}
We propose a quantum Fourier transform on photons in which a single atom-coupled cavity system mediates the photon-photon interactions. Our protocol utilizes time-delay feedback of photons and requires no active feedforward control. The time-delay feedback enables a single atom-cavity system to implement a quantum Fourier transform on an arbitrary number of photonic qubits on-the-fly, while rapid tuning of the atomic transition implements arbitrary controlled-phase gates. We analyze the performance of the protocol numerically and show that it can implement quantum Fourier transforms with tens of photons using state-of-the-art cavity quantum electrodynamics.
\end{abstract}

\maketitle

\section{Introduction}
% Put \section{\label{}} in argument of \section for cross-referencing
The quantum Fourier transform is a powerful tool in quantum information processing as it plays an essential role in many quantum algorithms, such as factoring~\cite{Shor2006} and phase estimation~\cite{Kitaev1995}. Additionally, it has applications in quantum simulation~\cite{Abrams1999, Kassal2011, Tichy2014}, quantum metrology~\cite{Humphreys2013, Su2017, Mukai2021, Vorobyov2021}, and quantum cryptography~\cite{Zhang2012, Yang2014}. Implementing quantum Fourier transforms was proposed in a variety of physical systems~\cite{Cirac1995, Fujiwara2005, Scully2002, Wang2011, Zhu2012, Mohamed2013}, and small-scale implementations were experimentally demonstrated using trapped ions~\cite{Gulde2003} and nuclear magnetic resonance~\cite{Chuang1998, Jones1998, Weinstein2001}. Extending this capability to photonics could significantly advance photonic quantum technologies and help take advantage of the photon's naturally weak interaction with the environment and facile distribution over optical fibers.

However, one of the challenges of implementing photonic quantum Fourier transforms is the lack of photon-photon interactions. Methods based on linear optics alone were proposed~\cite{Howell2000, Barak2007, Su2017, Kysela2020} and experimentally demonstrated~\cite{Bhattacharya2002}. However, these methods required exponentially increasing optical components for larger numbers of qubits. Other proposals using nonlinear optics exploited the weak cross-Kerr nonlinearity~\cite{Shen2013}, or alternately interactions between photons and two cavities coupled to individual quantum dots~\cite{Heo2019}. But they both required measurement and active feedforward control, in which one must apply unitary operations on the photons conditioned on the measurement results. Such feedforward adds significant overhead and typically requires efficient optical storage, which is challenging~\cite{Yamamoto2014, Jacobs2014}. As an alternative, coherent time-delay feedback control can eliminate measurements in the process and avoid introducing further decoherence~\cite{Carmele2013, Grimsmo2015, Carmele2016, Hein2016, Nemet2016, Guimond2016, Pichler2016, Guimond2017, Lu2017, Pichler2017, Whalen2017, Calajo2019, Nemet2019, Crowder2020, Zhan2020, Shi2021, Wan2021, ArranzRegidor2021, Barkemeyer2021}.

In this work, we propose a photon-based quantum Fourier transform where a single atom-coupled cavity system mediates the photon-photon interactions. Our approach does not require active feedforward control. Instead, it operates on-the-fly by taking a stream of sequential input photons and generating an output stream that contains the transforming result. The atom-cavity system mediates arbitrary conditional phase shifts between the photons by tuning the transition resonance of the atom. We analyze the performance of the protocol and provide a lower bound of the success probability using the diamond distance. We show that for the specific case of a quantum dot-cavity system, we can implement a quantum Fourier transform on tens of photons using state-of-the-art cavity quantum electrodynamics (cavity-QED).

\section{Protocol}
\begin{figure}[tb]
    \centering
    \includegraphics[width=1.0\columnwidth]{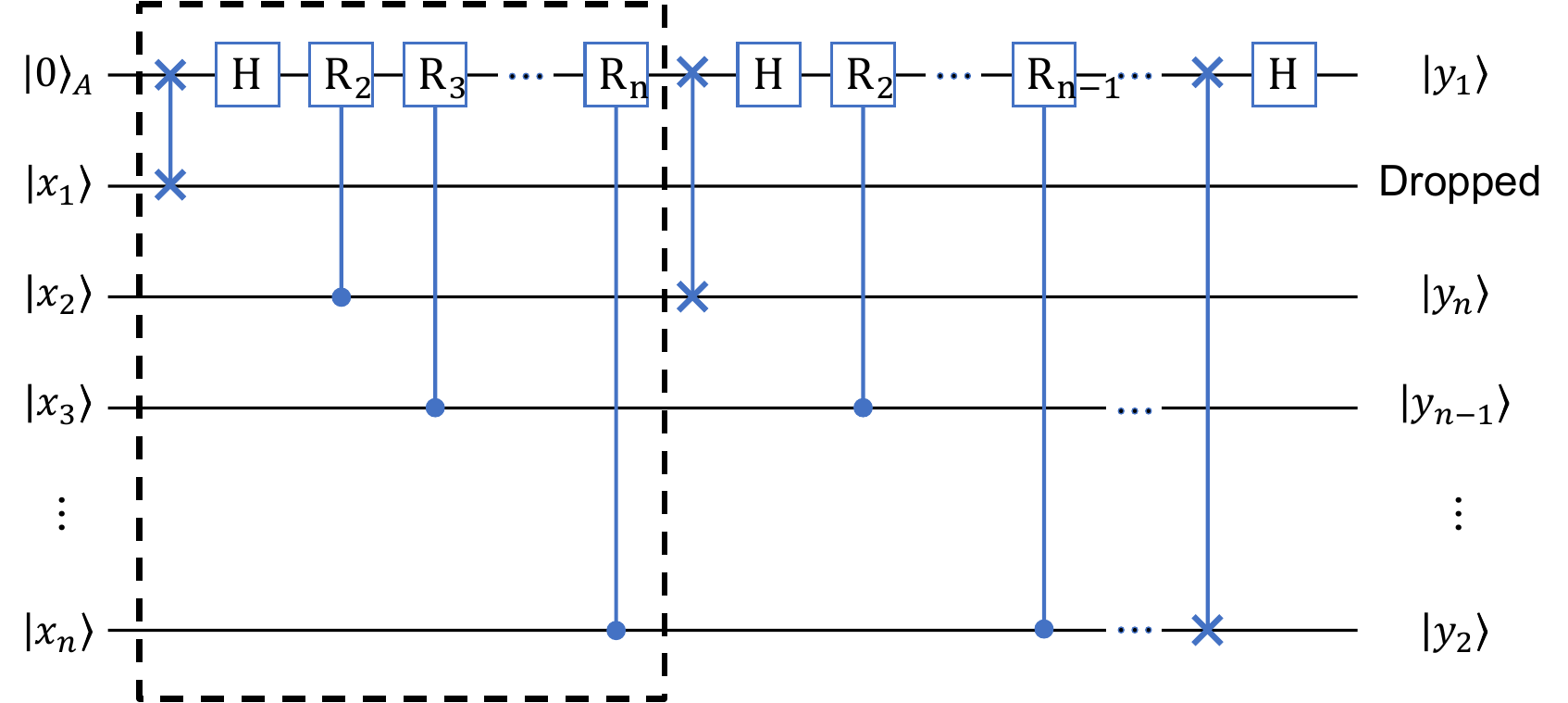}
    \caption{The circuit for the quantum Fourier transform on photonic qubits.}
    \label{fig:circuit}
\end{figure}

Our protocol implements the standard discrete quantum Fourier transform~\cite{Michael2010}, which requires a series of quantum phase gates between qubits. We utilize a single ancilla qubit to implement these phase gates, as illustrated by the quantum circuit in Figure~\ref{fig:circuit}. This circuit takes n photonic input qubits and performs a quantum Fourier transform (QFT)
\[\left|x_1x_2\cdots x_n\right\rangle\xrightarrow{\rm{QFT}}\frac{1}{2^{n/2}}\sum_{y=1}^{2^n-1}{e^{i2\pi\cdot x\cdot y/2^n}\left|y_1y_2\cdots y_n\right\rangle}\;,\]
where $n$ is the total number of input photons, and $x=x_1x_2\cdots x_n$ and $y=y_1y_2\cdots y_n$ are the binary representations of the input and output qubits, respectively. The vertical line with crossed ends represents a swap gate, which exchanges states between the atom and photons. The box labeled $H$ represents a Hadamard gate. The box labeled $R_k$ with the photonic qubit connected via the vertical line represents a controlled-phase gate by the matrix
\begin{equation}
    CR_k=\begin{pmatrix}1&0&0&0\\0&1&0&0\\0&0&1&0\\0&0&0&e^{i2\pi/2^k}\end{pmatrix}\;,
\end{equation}
where $k=2,3,\cdots,n$. After implementing the transform, we can drop the first photon because it carries the initial state of the ancilla.

The circuit in Figure~\ref{fig:circuit} consists of $n$ subroutines. In the first subroutine, indicated by the dashed box in the figure, the circuit implements a swap gate between the ancilla and the first photon, which subsequently leaves the system. The ancilla then applies a $CR_k$ gate to the remaining photonic qubits $2$ through $n$. In the second subroutine, we repeat the procedure for the second photonic qubit, in which we apply a swap gate and photon $2$ leaves the system. The ancilla then applies a $CR_k$ gate to photons $3$ to $n$. We repeat the procedure until all photons have left the system. Once the protocol completes, the atomic qubit contains the first qubit of the Fourier transform. We can directly measure it from the atom or alternately inject another photon into the system and implement a swap to move the state back to an all-photonic Hilbert space.

\begin{figure}[tb]
    \centering
    \includegraphics[width=1.0\columnwidth]{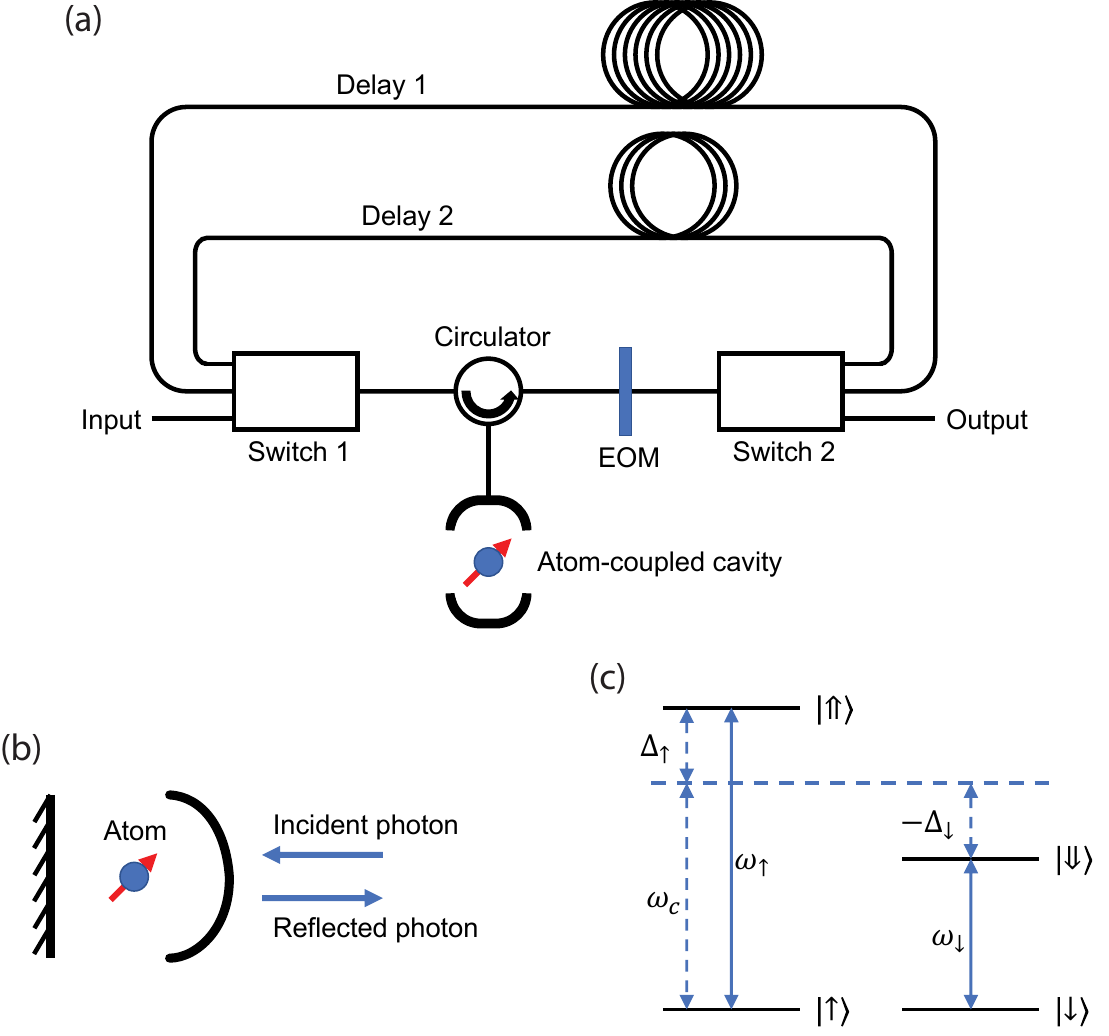}
    \caption{(a) The schematic setup for a photonic quantum Fourier transform. (b) The atom-coupled single-sided cavity system. (c) The level structure of the atom-cavity system.}
    \label{fig:setup}
\end{figure}

To implement the circuit in Figure~\ref{fig:circuit}, we propose the optical setup shown in Figure~\ref{fig:setup}(a). The ancilla qubit is a single atom coupled to an optical cavity. The cavity mediates strong interactions with the photons that generate the $CR_k$ gates, as discussed in more detail below. The photons are injected into the system sequentially, separated by a time delay of $T_{\rm{cycle}}$ (operation cycle). To implement the sequence of subroutines, we utilize two time-delay feedback loops and two switches. Delay line $1$ generates a delay of $\tau_1>nT_{\rm{cycle}}$. This delay stores the remaining qubits after the execution of each subroutine, which will be used in the next subroutine. Delay line $2$ generates a delay of $\tau_2\ll T_{\rm{cycle}}$. This short delay enables a single photon to interact with the atomic qubit multiple times, which is necessary to create a swap gate.

The protocol starts by injecting all $n$ photons at the input. To implement the swap gate, we relay and reflect photon $1$ off the cavity three times through delay line $2$. Each reflection implements a $CR_1$ gate between the photon and the atom. The three $CR_1$ gates combined with the Hadamard gates implement the swap gate between the photon and the atom as given by the identity
\begin{equation}
    \mathit{SWAP}=H_{a,p}\cdot CR_1\cdot H_{a,p}\cdot CR_1\cdot H_{a,p}\cdot CR_1\;,
\end{equation}
where $H_{a,p}=H_a\otimes H_p$ represents a Hadamard gate both on the atom~\cite{Press2008, Carter2013} and photon~\cite{OBrien2007}. After the reflections, photon $1$ exits the system from the output. The remaining $n-1$ photons then sequentially reflect off the cavity, which implement the controlled-phase gates, and enter delay line $1$, which transmits them back to the input for the second subroutine. We repeat the procedure until all photons leave the system. 

We next describe how the atom-cavity system implements the $CR_k$ gates. We consider an atom in a single-sided cavity, as shown in Figure~\ref{fig:setup}(b). This system is already known to generate the controlled-phase flip ($CR_1$) gate between the atomic and photonic qubits using cavity-mediated interactions~\cite{Duan2004}. Furthermore, we will propose a method to fast apply any $CR_k$ gates on sequential photons by active phase tuning. This method also solves the problem of requiring extremely high magnetic field to implement the $CR_1$ gate in experiments~\cite{Sun2014, Shi2021}.

The system features a level structure as shown in Figure~\ref{fig:setup}(c). The atom possesses two ground states ($\left|\uparrow\right\rangle$, $\left|\downarrow\right\rangle$) and two excited states ($\left|\Uparrow\right\rangle$, $\left|\Downarrow\right\rangle$), in which the quantization axis is along the direction of an externally applied magnetic field that breaks the degeneracy of the levels. The cavity only supports a linear polarization mode along the quantization axis, which couples to the transition between $\left|\uparrow\right\rangle$ and $\left|\Uparrow\right\rangle$, and the transition between $\left|\downarrow\right\rangle$ and $\left|\Downarrow\right\rangle$. We define this mode as the vertical polarization mode and its orthogonal polarization as the horizontal polarization mode.  We denote the frequency of the cavity mode as $\omega_c$, the spin-up atomic transition frequency as $\omega_\uparrow$, and the spin-down atomic transition frequency as $\omega_\downarrow$. We also denote the detuning between the atomic transition and the cavity mode as $\Delta_{\uparrow,\downarrow}=\omega_{\uparrow,\downarrow}-\omega_c$.

The Hamiltonian of the atom-cavity system is given by
\begin{eqnarray}
    \mathcal{H}/\hbar&=&\omega_c{\hat{a}}^\dag\hat{a}+\omega_\uparrow\sigma_{\Uparrow\Uparrow}+
    \omega_\downarrow\sigma_{\Downarrow\Downarrow}\nonumber\\
    & &+ig\left(\sigma_{\Uparrow\uparrow}\hat{a}-\sigma_{\uparrow\Uparrow}{\hat{a}}^\dag+
    \sigma_{\Downarrow\downarrow}\hat{a}-\sigma_{\downarrow\Downarrow}{\hat{a}}^\dag\right)\;,
\end{eqnarray}
where $\hat{a}$ is the operator for the cavity mode in vertical polarization, $\sigma$ is the atomic operator, and $g$ is the atom-cavity coupling strength. The atomic state space is reducible into two uncoupled subspaces $\left\{\left|\uparrow\right\rangle,\ \left|\Uparrow\right\rangle\right\}$ and $\left\{\left|\downarrow\right\rangle,\ \left|\Downarrow\right\rangle\right\}$. In either subspace, we can write the Heisenberg equations of motion for the atom-cavity system and the external field as~\cite{Walls2007}
\begin{eqnarray}
    \frac{d\hat{a}}{dt}&=&-\left[i\left(\omega_c-\omega\right)+\frac{\gamma}{2}\right]\hat{a}-g\sigma_--\sqrt{\gamma}{\hat{a}}_{\rm in}\;,\nonumber\\
    \frac{d\sigma_-}{dt}&=&-\left[i\left(\omega_{\uparrow,\downarrow}-\omega\right)+\frac{\kappa}{2}\right]\sigma_--g\sigma_z\hat{a}\;,\nonumber\\
    {\hat{a}}_{\rm out}&=&{\hat{a}}_{\rm in}+\sqrt{\gamma}\hat{a}\;,
\end{eqnarray}
where ${\hat{a}}_{\rm in}$ and ${\hat{a}}_{\rm out}$ are the input and output field operators (probe beam), respectively, and $\omega$ is the frequency of the probe beam, which we set as $\omega=\omega_c$. The parameter $\kappa$ is the atom dipole decay rate, and $\gamma$ is the cavity decay rate. We can solve the equation of motion for a single-photon input in the quasi-monochromatic limit. The probe beam experiences a state-dependent phase shift. When the probe beam is in the horizontal polarization (H), it does not couple to the cavity but reflects from a mirror and acquires no phase shift. However, when the probe beam is in the vertical polarization (V), it reflects off the cavity, where the reflection coefficient is given by
\begin{equation}
    r_{\uparrow.\downarrow}=\frac{C_{\uparrow,\downarrow}-1}{C_{\uparrow,\downarrow}+1}\;.
\end{equation}
In the above equation $C_{\uparrow,\downarrow}=\frac{4g^2}{\gamma\left(\kappa+i2\Delta_{\uparrow,\downarrow}\right)}$ is the spin-dependent cooperativity. We define on-resonant cooperativity $C=\frac{4g^2}{\gamma\kappa}$ such that the relation between the on-resonant and spin-dependent cooperativities is given by $C_{\uparrow,\downarrow}=\frac{C}{1+2i\Delta_{\uparrow,\downarrow}/\kappa}$. In the high cooperativity of $C\gg1$, the reflection coefficient is given by $r_{\uparrow,\downarrow}=e^{i\theta_{\uparrow,\downarrow}}$, where
\begin{equation}
    \theta_{\uparrow,\downarrow}=\mathrm{Im}\left\{\ln{\frac{1-i2\Delta_{\uparrow,\downarrow}/\kappa C}{1+i2\Delta_{\uparrow,\downarrow}/\kappa C}}\right\}\;,
\end{equation}
which is continuously tunable from $0$ to $2\pi$ by changing the detuning $\Delta_{\uparrow,\downarrow}$~\cite{Sun2016}. After the reflection, we apply a single-qubit phase gate of $\begin{pmatrix}1&0\\0&e^{-i\theta_\uparrow}\end{pmatrix}$ on the photon. The reflection and the photonic phase gate implement a controlled-phase shift gate on the photon and atom given by
\begin{equation}
    U_{\rm cp}=\begin{pmatrix}1&0&0&0\\0&1&0&0\\0&0&1&0\\0&0&0&e^{i\Delta\theta}\end{pmatrix}\;,
\end{equation}
where $\Delta\theta=\theta_\downarrow-\theta_\uparrow$ is the relative phase shift. In the above operators, we denote the state of the input photon as $\left|H\right\rangle\equiv\left|0\right\rangle_p$ for the horizontal polarization and $\left|V\right\rangle\equiv\left|1\right\rangle_p$ for the vertical polarization, the atomic state as $\left|\uparrow\right\rangle\equiv\left|0\right\rangle_a$ and $\left|\downarrow\right\rangle\equiv\left|1\right\rangle_a$, and the composite state space as $\left|x\right\rangle_p\otimes\left|y\right\rangle_a$. 

The setup illustrated in Figure~\ref{fig:setup}(a) requires the ability to apply different $CR_k$ gates to sequentially reflected photons. To do so requires the ability to modulate the detunings $\Delta_\uparrow$ and $\Delta_\downarrow$ rapidly. These detunings are given by $\Delta_\uparrow=\omega_\uparrow-\omega_c$ and $\Delta_\downarrow=\Delta_\uparrow+\Delta_Z$, where $\Delta_Z$ is the Zeeman splitting of the spin transitions. The Zeeman splitting is strictly determined by the applied magnetic field, which is difficult to modulate rapidly. Alternatively, we apply a constant Zeeman splitting and focus on the active tuning of the atomic transition $\omega_\uparrow$. In particular, both the quantum-confined Stark effect~\cite{Faraon2010, White2020} and AC Stark shift~\cite{Bose2014} can modulate the detuning on nanosecond timescales, which allows us to change the phase of the $CR_k$ gates quickly.

\section{Active phase gate tuning\label{sec:tuning}}
To calculate the achievable phase shifts by tuning the atomic transition, we consider the specific case of a charged quantum dot coupled to a nanophotonic cavity~\cite{Hennessy2007, Carter2013, Schaibley2013, Sun2016}. We denote the spin-dependent detuning as $\Delta_\uparrow=\Delta_S+\Delta_0$ and $\Delta_\downarrow=\Delta_S+\Delta_Z+\Delta_0$ under the Stark shift and Zeeman splitting, where $\Delta_S$ is the Stark shift and $\Delta_0$ is an offset. We set $\Delta_0=\frac{\kappa}{2}\sqrt{C^2-1}$ and $\Delta_Z=-2\Delta_0$, such that the controlled-phase shift $\Delta\theta=\pi$ when $\Delta_S=0$. 
The Zeeman splitting is determined by the magnetic field, as $\Delta_Z=\left(g_e+g_h\right)\mu_BB/\hbar$, where $g_e$ and $g_h$ are Lande factors for the electron and hole, $\mu_B$ is the Bohr magneton, and $B$ is the applied magnetic field. 
For this quantum dot-cavity system, we can achieve the parameters: $g_e=0.43$, $g_h=0.21$, $g=11\ {\rm GHz}$, $\kappa=0.3\ {\rm GHz}$, and $\gamma=28\ {\rm GHz}$~\cite{Sun2014}.

\begin{figure}[tb]
    \centering
    \includegraphics[width=1.0\columnwidth]{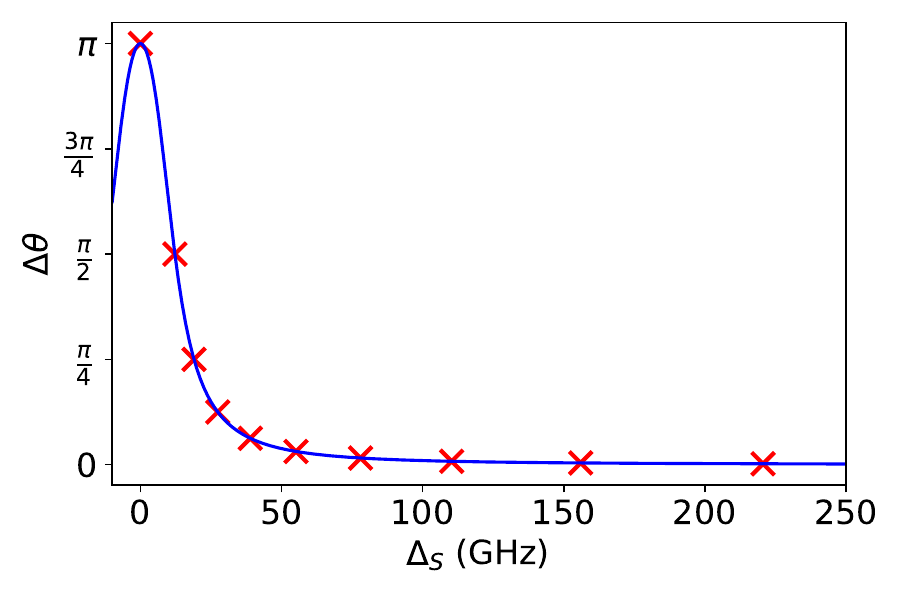}
    \caption{The controlled-phase shift $\Delta\theta$ as a function of the Stark shift $\Delta_S$.}
    \label{fig:tuning}
\end{figure}

Figure~\ref{fig:tuning} shows the controlled-phase shift $\Delta\theta$ as a function of the Stark shift $\Delta_S$. We set the offset detuning to $\Delta_0=8.64\ {\rm GHz}$ and apply a magnetic field of $B=1.93\ {\rm T}$. The red marks indicate the specific Stark shift used to implement $CR_k$ gates for $k=1,2,\cdots,10$. Given large $k$, the corresponding Stark shift scales as $\Delta_S\sim\kappa C\sqrt{\frac{2^k}{2\pi}}$. Therefore, implementing a small phase shift of $\frac{2\pi}{2^k}$ requires a high detuning, which may go beyond practical values. But we can discard the $CR_k$ gates when $k$ is larger than some cutoff $K$ in the Fourier transform circuit. Those gates are close to an identity operation exponentially, so discarding them would raise little error~\cite{Hales2000}. The state-of-the-art Stark shift is achievable at $\sim1000\ {\rm GHz}$ for quantum dots~\cite{Aghaeimeibodi2019}. This tuning enables the $CR_k$ gates with a maximum $k$ of $14$, which is sufficiently large for a highly precise quantum Fourier transform.

\section{Analysis}
To analyze the performance of the protocol under realistic conditions, we use the diamond distance~\cite{Aharonov1998, Gilchrist2005, Watrous2009, Shi2021a}. The diamond distance between two quantum channels $\mathcal{E}$ and $\mathcal{F}$ on a n-dimensional state space is defined as
\begin{equation}
    d_\diamond\left(\mathcal{E},\mathcal{F}\right)=\frac{1}{2}\max_{\rho}{\left\|\left(\mathcal{E\otimes I}\right)\rho-\left(\mathcal{F\otimes I}\right)\rho\right\|_1}\;,
\end{equation}
where $\left\|\cdot\right\|_1$ denotes the trace norm, $\mathcal{I}$ represents an identity channel on the auxiliary space, and $\rho$ is a quantum state on the extended space. One property of the diamond distance is chaining
\begin{equation}
    d_\diamond\left(\mathcal{E}_2\circ\mathcal{E}_1,\mathcal{F}_2\circ\mathcal{F}_1\right)\le d_\diamond\left(\mathcal{E}_1,\mathcal{F}_1\right)+d_\diamond\left(\mathcal{E}_2,\mathcal{F}_2\right)\;,
\end{equation}
where $\mathcal{E}_2\circ\mathcal{E}_1$ represents a process of $\mathcal{E}_1$ followed by $\mathcal{E}_2$. Using the chaining property, we can calculate the distance of a whole process by summing the distance of its individual steps. To characterize the error of the photonic quantum Fourier transform, we can calculate the diamond distance between the non-ideal version implemented in Figure~\ref{fig:setup} and the ideal circuit. The diamond distance gives an upper bound probability of error of the output~\cite{Gilchrist2005}. We define the success probability as
\begin{equation}
    P_s=1-\mathcal{D}\;,
\end{equation}
where $\mathcal{D}$ is the diamond distance between the implemented and ideal transforms. $P_s$ gives a lower bound probability of the correct output.

Using the chaining property, we can express the diamond distance as
\begin{eqnarray}
    \mathcal{D}&=&N^2d_p+2Nd_H+3Nd_1+\sum_{k=2}^{K}{\left(N-k+1\right)d_k}\nonumber\\
    & &+\sum_{k=K+1}^{N}{\left(N-k+1\right)d_k^\star}\;,
    \label{eq:distance}
\end{eqnarray}
where $N$ is the number of photonic qubits. Each term represents the diamond distance of different channels, whose coefficient counts the implementation of the channel in the circuit. The term $d_p$ is the distance between the atomic qubit dephasing channel and identity channels in one operation cycle $T_{\rm cycle}$; $d_H$ is the distance between the imperfect and ideal atomic Hadamard gates; $K$ is the cutoff for $k$ as discussed in Section~\ref{sec:tuning}; $d_k$ is the distance between the operation of reflecting a photon off the atom-coupled cavity and an ideal $CR_k$ gate; $d_k^\star$ is the distance between the $CR_k$ and identity gates. We can calculate each term through convex optimization~\cite{Watrous2009}, which leads to the expressions
\begin{eqnarray}
    d_p&=&\frac{1}{2}\left(1-e^{-T_{cycle}/T_2}\right)\;,\nonumber\\
    d_H&=&p\;,\nonumber\\
    d_k^\star&=&\frac{1}{2}\left|1-e^{i2\pi/2^k}\right|\;.
\end{eqnarray}
In the above equations, $T_2$ is the characteristic dephasing time of the atom, $p$ is the error probability of atomic Hadamard gate.

The one distance which involves subtleties is $d_k$, the distance between the non-ideal $CR_k$, implemented by the atom-cavity system, and the ideal $CR_K$ gate. This channel features a state-dependent photon loss which is not a trace-preserving operation. To properly define the distance, we can model the imperfect reflection as an ideal $CR_k$ gate followed by the measurement and post-selection operation
\begin{equation}
    \mathcal{M}\left(\rho\right)=\frac{M\rho M^\dag}{Tr\left[M\rho M^\dag\right]}\;,
\end{equation}
where $M$ is a measurement operator as
\begin{equation}
    M=\begin{pmatrix}1&0&0&0\\0&1&0&0\\0&0&\left|r_\uparrow\right|&0\\0&0&0&\left|r_\downarrow\right|\end{pmatrix}\;.
\end{equation}
Thus, $d_k$ is given by the distance between the post-selection operation and the identity channel. Unfortunately, the post-selection channel applies a nonlinear transform on the photon-atom state so that we cannot calculate its distance by convex optimization. In Appendix~\ref{app:distance}, we show that in this post-selected case, we can calculate the distance using the maximum and minimum eigenvalues of $M$, such that
\begin{equation}
    d_k=\frac{1-\min{\left(\left|r_\uparrow\right|,\left|r_\downarrow\right|\right)}}{1+\min{\left(\left|r_\uparrow\right|,\left|r_\downarrow\right|\right)}}\;.
\end{equation}
It approximates $\frac{1}{2C^2+8\Delta^2/\kappa^2}$ for the atom-coupled cavity system, where $\Delta=\min{\left(\Delta_\uparrow,\Delta_\downarrow\right)}$.

State-independent photon loss is another type of error that can occur in optical systems during the process of optical transmission and detection. Such loss decreases the overall success probability but does not affect the fidelity of the post-selected output states. A particular source of loss is the delay in the protocol. The $k$-th photon passes a total delay length of $(k-1)cNT_{\rm cycle}$ in the quantum Fourier transform of $N$ photons, where $c$ is the speed of light. We can beat photon loss by repeating the protocols and post-selection or by using redundant encoding~\cite{Ralph2005, Lu2008, Bergmann2016}. In the following simulation, we calculate the post-selected success probability (a measure of fidelity) of the protocol.

\begin{figure}[tb]
    \centering
    \includegraphics[width=1.0\columnwidth]{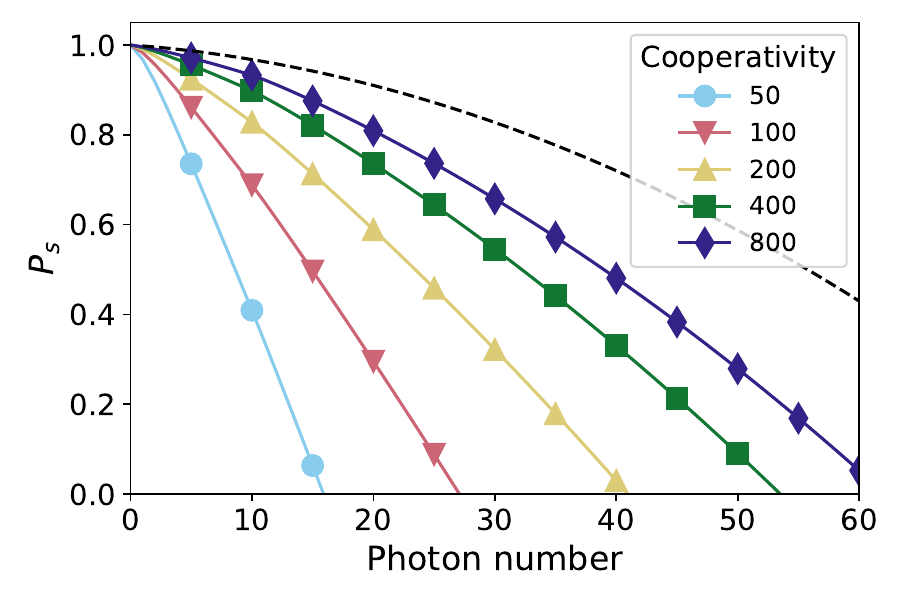}
    \caption{The success probability $P_s$ as a function of photon number under different cooperativities.}
    \label{fig:success}
\end{figure}

Having attained an expression for all terms in the diamond distance, we can now analyze a photonic quantum Fourier transform under realistic experimental conditions. We use the same parameters as those in Section~\ref{sec:tuning} for a charged quantum dot in a cavity and set an operation cycle of $T_{\rm cycle}=5\ {\rm ns}$. Figure~\ref{fig:success} shows the success probability $P_s$ as a function of photon number $N$, where we assume a spin characteristic dephasing time of $20\ {\rm \mu s}$ and a spin Hadamard gate error rate of $0.001$. The curves indicated by markers show the success probability under finite cooperativities, where we set $K=10$ in Equation~\ref{eq:distance} as the largest implemented $CR_k$ gates. The dashed black line shows the success probability for an ideal $CR_k$ gate for comparison. The success probability decreases quadratically as a function of the photon number for all cooperativities. Higher cooperativity can increase $P_s$, but this improvement becomes marginal when the cooperativity exceeds $400$. In this limit, $P_s$ is mainly determined by the spin error. A valid threshold for $P_s$ depends on the application of the quantum Fourier transform. For example, in the case of factoring~\cite{Shor2006}, which is a verifiable algorithm, any non-zero $P_s$ can guarantee the effectiveness of the transform. We must repeat the experiment for $O\left(1/P_s\right)$ times before we can get a correct result. Thus, we can achieve tens of photonic qubits quantum Fourier transform using state-of-the-art devices of $\sim200$ cooperativity. However, other applications in the quantum information process may require a higher success probability.

\begin{figure}[tb]
    \centering
    \includegraphics[width=1.0\columnwidth]{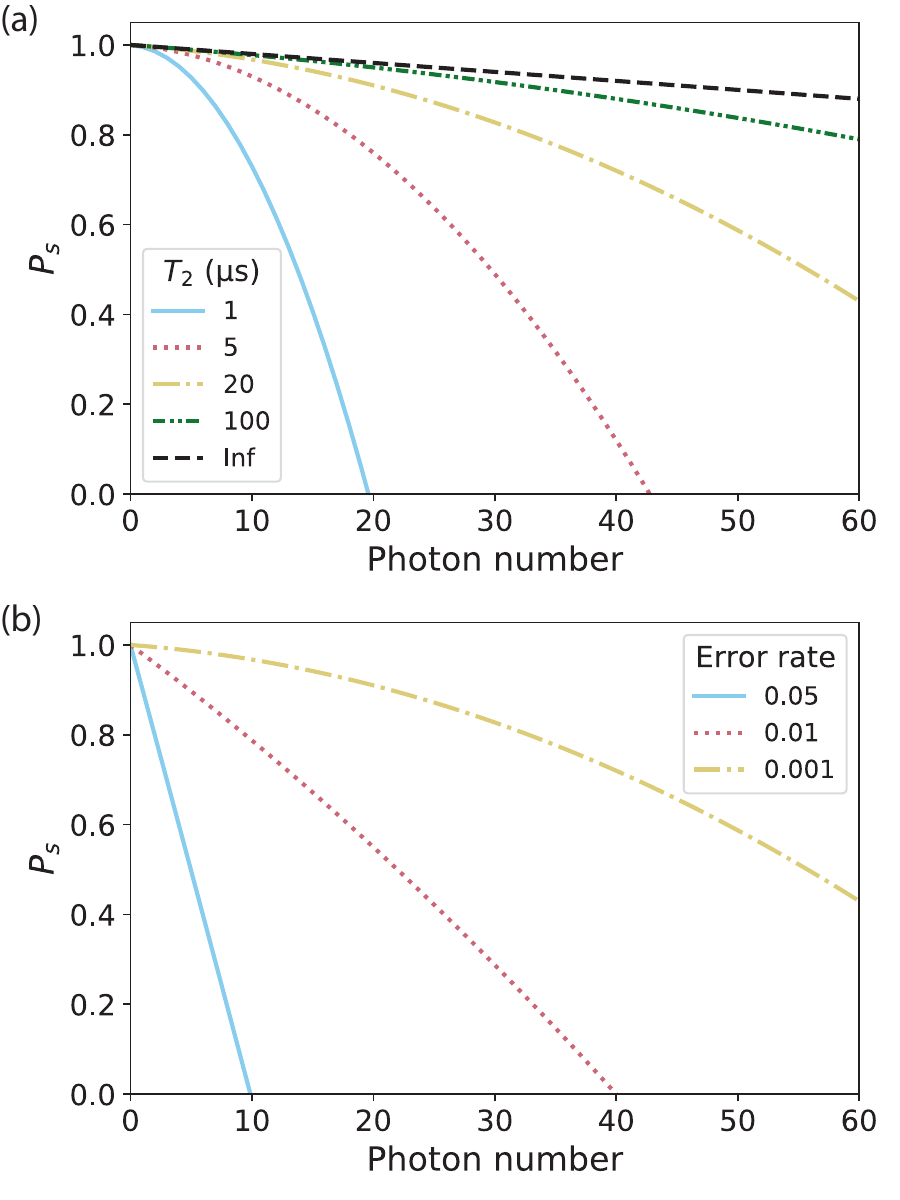}
    \caption{The success probability $P_s$ as a function of photon number, (a) under different spin dephasing time $T_2$ assuming the spin Hadamard gate error rate of $0.001$ and ideal $CR_k$ gates, (b) under different spin Hadamard gate error rates assuming $T_2=20\ {\rm \mu s}$ and ideal $CR_k$ gates.}
    \label{fig:spin}
\end{figure}

To investigate the effect of spin errors, we calculate the success probability under different dephasing time $T_2$, as shown in Figure~\ref{fig:spin}(a), where we assume ideal $CR_k$ gates and set $p=0.001$. We define the maximum number of photons that the protocol can process as the highest value for $N$ for which our bound is non-zero. For $T_2=1\ {\rm \mu s}$ , the protocol can process about $20$ photons. Increasing the dephasing time from $1\ {\rm \mu s}$ to $100\ {\rm \mu s}$ can improve the success probability. However, as $T_2$ ranges from $100\ {\mu s}$  to infinity, increasing $T_2$ hardly changes $P_s$ for quantum Fourier transform of tens of photons.

Figure~\ref{fig:spin}(b) shows the success probability under different spin Hadamard gate error rates, assuming the dephasing time $T_2=20\ {\rm \mu s}$ and ideal $CR_k$ gates. When $p=0.05$, the protocol can process a very limited number of photons before the bound of the success probability goes to zero. Otherwise, when $p=0.001$, the success probability is largely improved. Considering a spin qubit in a quantum dot with a dephasing time of microseconds is currently achievable~\cite{Press2008}, the bottleneck of experimentally implementing the quantum Fourier transform within tens of photons is the optimization of the quantum control of the spin rotation. This parameter is ultimately the main limitation for the current implementation of the protocol.

A related protocol~\cite{Heo2019} also implements quantum Fourier transform via quantum dot-coupled cavities, which applies a controlled-phase gate by three reflections without dynamic tuning of the quantum dot but requires two measurements and feed-forward operations. It is a total of $\frac{3}{2}N(N-1)$ reflections and $N(N-1)$ measurement-based operations for an $N$-qubit Fourier transform. In comparison, our protocol uses $\frac{1}{2}N(N+5)$ reflections, which relaxes the requirement for the quantum dot-cavity cooperativity and removes the need for the challenging measurement and feed-forward control~\cite{Yamamoto2014, Jacobs2014}. In addition, the measurement-based control is no faster than active phase tuning in our protocol in terms of processing time. For example, measuring a spin qubit in a quantum dot takes $3.4\ {\rm ns}$ on average~\cite{Press2008}, and the succeeding classical control requires additional time.

\section{Conclusion}
We propose a photonic quantum Fourier transform protocol, which uses a single atom-coupled cavity system to implement photon-photon interactions and requires no active feedforward control. We show that a quantum Fourier transform with tens of photons may be possible with state-of-the-art cavity-QED systems. For future study, we expect to implement other quantum algorithms, such as the variational quantum eigensolver~\cite{Peruzzo2014}, using the same on-the-fly system. Ultimately, the proposed quantum Fourier transform could enable fast on-the-fly photonic quantum information processing, which may be particularly useful in compact quantum photonic circuits where active feedforward is challenging.

% % If you have acknowledgments, this puts in the proper section head.
\begin{acknowledgments}
The authors would like to acknowledge financial support from the National Science Foundation (grant number OMA1936314 and ECCS1933546), the Air Force Office of Scientific Research (grant number UWSC12985 and FA23862014072), and the Army Research Laboratory (grant W911NF1920181).
\end{acknowledgments}

\appendix

\section{Distance for the measurement and post-selection operation\label{app:distance}}
We calculate the diamond distance between the measurement and post-selection operation and the identity channel, as
\begin{equation}
    d\left(\mathcal{I},\mathcal{M}\right)=\frac{1}{2}\max_{\rho}{\left\|\rho-\frac{M\rho M^\dag}{Tr\left[M\rho M^\dag\right]}\right\|_1}\;,
\end{equation}
where $M$ is a nonnegative diagonal matrix. Readers can refer to~\cite{Shi2021a} for a more detailed discussion, which considers the distance for any non-trace preserving operations. The distance is maximized on a pure state $\psi$. Thus, we can calculate it by minimizing $\cos{\theta}=\left|\left\langle\psi\middle|\phi\right\rangle\right|$ and using the identity $\frac{1}{2}\left\|\psi\psi^\dag-\phi\phi^\dag\right\|_1=\sqrt{1-\left|\left\langle\psi\middle|\phi\right\rangle\right|^2}$, where $\left|\phi\right\rangle=\frac{M\left|\psi\right\rangle}{\left|M\left|\psi\right\rangle\right|}$ is the normalized state after measurement.

To calculate $\cos{\theta}$, we consider a diagonal matrix $M_n$ in an $n$-dimensional state space, where $M_n=diag\left(\lambda_1,\lambda_2,\cdots,\lambda_n\right)$ and $\lambda_1\geq\lambda_2\geq\cdots\lambda_n\geq0$. We denote a unit vector $\left|u\right\rangle=\sqrt{x}\left|n\right\rangle+\sqrt{1-x}\left|v\right\rangle$, where $x\in\left[0,1\right]$, $\left|n\right\rangle$ is the basis corresponding to the eigenvalue $\lambda_n$, and $\left|v\right\rangle$ is a unit vector in the $\left(n-1\right)$-dimensional subspace orthogonal to $\left|n\right\rangle$. The assumption of positive coefficients loses no generality because we can always redefine the basis with a phase factor. Applying $M_n$ to $\left|u\right\rangle$ gives
\begin{equation}
    M_n\left|u\right\rangle=\lambda_n\sqrt{x}\left|n\right\rangle+\sqrt{1-x}M_{n-1}\left|v\right\rangle\;,
\end{equation}
where $M_{n-1}$ is a diagonal matrix in the subspace. Applying $M_{n-1}$ to $\left|v\right\rangle$ will transform the vector as
\begin{equation}
    M_{n-1}\left|v\right\rangle=r_{n-1}\cos{\theta_{n-1}}\left|v\right\rangle+r_{n-1}\sin{\theta_{n-1}}\left|v_\perp\right\rangle\;,
\end{equation}
where $r_{n-1}$ and $\theta_{n-1}$ are the contracting factor and rotation angle, and $\left|v_\perp\right\rangle$ is a unit vector orthogonal to $\left|v\right\rangle$. A specific rotating axis does not affect the following calculation because the axis is always orthogonal to $\left|n\right\rangle$. We calculate the contracting factor $r_n=\sqrt{\left|\left\langle u\middle|M_n^2\middle|u\right\rangle\right|}$ and the rotating angle $\cos{\theta_n}=\left\langle u\middle|M_n\middle|u\right\rangle/r_n$ for applying $M_n$ to $\left|u\right\rangle$ and get
\begin{eqnarray}
    r_n^2&=&x\lambda_n^2+(1-x)r_{n-1}^2\;,\nonumber\\
    r_n\cos{\theta_n}&=&x\lambda_n+(1-x)r_{n-1}\cos{\theta_{n-1}}\;,
\end{eqnarray}
i.e., the feasible point $\left(r_n^2,r_n\cos{\theta_n}\right)$ is a convex combination of $\left(\lambda_n^2,\lambda_n\right)$ and $\left(r_{n-1}^2,r_{n-1}\cos{\theta_{n-1}}\right)$. By induction, we can compute the feasible points of $\left(r^2,r\cos{\theta}\right)$ for applying $M$ to a pure state, as
\begin{eqnarray}
    r^2&=&\sum_{i}{p_i\lambda_i^2}\;,\nonumber\\
    r\cos{\theta}&=&\sum_{i}{p_i\lambda_i}\;,
\end{eqnarray}
where $\sum_{i}{p_i}=1$. The minimum of $\cos{\theta}$ is on the boundary of the feasible domain and equals $\frac{2\sqrt{\lambda_1\lambda_n}}{\lambda_1+\lambda_n}$, where $\lambda_1$ and $\lambda_n$ are the maximum and minimum eigenvalues of $M$. Therefore, the distance between a measurement channel $\mathcal{M}$ and an identity channel is given by
\begin{equation}
    d\left(\mathcal{I},\mathcal{M}\right)=\frac{\lambda_1-\lambda_n}{\lambda_1+\lambda_n}\;.
\end{equation}

% \bibliography{main.bib}
% \bibliographystyle{apsrev4-2}

\end{document}